\documentclass{article}
\PassOptionsToPackage{numbers, compress}{natbib}
\usepackage[final]{neurips_2019}
\usepackage{graphicx}
\graphicspath{ {./images/} }
\usepackage[utf8]{inputenc} 
\usepackage[T1]{fontenc}    
\usepackage{hyperref}       
\usepackage{url}            
\usepackage{booktabs}       
\usepackage{amsfonts}       
\usepackage{nicefrac}       
\usepackage{microtype}      
\usepackage{xcolor}         
 \usepackage{multirow}

\title{Intelligent Text-Conditioned Music Generation}

\author{
    Zhouyao Xie\\
    Language Technologies Institute\\
    Carnegie Mellon University\\
    Pittsburgh, PA 15213 \\
    {zhouyaox@cs.cmu.edu}
    \And
    Nikhil Yadala\\
    Language Technologies Institute\\
    Carnegie Mellon University\\
    Pittsburgh, PA 15213 \\
    {nyadala@alumni.cmu.edu}
    \And
    Xinyi Chen\\
    Language Technologies Institute\\
    Carnegie Mellon University\\
    Pittsburgh, PA 15213 \\
    {xinyic@alumni.cmu.edu}
    \And
    Jing Xi Liu\\
    Language Technologies Institute\\
    Carnegie Mellon University\\
    Pittsburgh, PA 15213 \\
    {jingxil@alumni.cmu.edu}
    \And
    Eric Nyberg\\
    Language Technologies Institute\\
    Carnegie Mellon University\\
    Pittsburgh, PA 15213 \\
    {ehn@alumni.cmu.edu}
}

\begin{document}
\maketitle

\begin{abstract}
    CLIP (Contrastive Language-Image Pre-Training) \cite{Radford2021LearningTV} is a multimodal neural network trained on (text, image) pairs to predict the most relevant text caption given an image. It has been used extensively in image generation by connecting its output with a generative model such as VQGAN, with the most notable example being OpenAI's DALLE-2 \cite{dalle2}. In this project, we apply a similar approach to bridge the gap between natural language and music. Our model is split into two steps: first, we train a CLIP-like model on pairs of text and music over contrastive loss to align a piece of music with its most probable text caption. Then, we combine the alignment model with a music decoder to generate music. To the best of our knowledge, this is the first attempt at text-conditioned deep music generation. Our experiments show that it is possible to train the text-music alignment model using contrastive loss and train a decoder to generate music from text prompts.
\end{abstract}
\vskip 1ex
\section{Introduction}

Our project has been inspired by recent advancements in multimodal machine learning and deep generative models, including but not limited to CLIP \cite{Radford2021LearningTV}, DALLE-2 \cite{dalle2}, MelGAN \cite{kumar2019}, and MusicVAE \cite{pmlr-v80-roberts18a}. Specifically, our project focuses on the task of deep music generation, i.e. the task of generating music compositions using deep learning.
Recent advances in deep music generation include using deep learning architectures such as WaveNet and Long-Short Term Memory (LSTM) to generate music \cite{kumar2019}, but these approaches are flawed in either capturing long-term signal patterns or sparse conditional signals. In addition, there has been recent works done in domains such as text-to-image or text-to-audio models that bridged the gap between advancements in text and image modeling \cite{sReed2016, ramesh2021}. These findings exploit a potential alternative that makes use of a considerably larger stream of supervision for scalable multi-model representations \cite{Radford2021LearningTV}. However, very little work has been done in the music field. Major reasons include a lack of large multimodal music datasets, a lack of effective and unified evaluation metrics, as well as the music itself being more challenging than image and audio. 

In this research-oriented project, we aim to address these challenges by proposing an intelligent text-conditioned music generation framework, as well as a set of evaluation metrics on generated music based on three aspects, integrity, pitch, and rhythm. We propose the possibility of training a decoder to generate music from natural language prompts and an encoder that is capable of learning music representation from text inputs. We hypothesize that such approaches will yield better results than unimodal generators.

\section{Hypothesis/Project Goal}

We hypothesize that
\begin{enumerate}
    \item It is possible to learn a mutual latent space for music and text in which matching music and text pairs are close to each other using contrastive learning and cross-attention;
    \item It is possible to perform conditional generation from the learned latent space by training a music decoder;
    \item It is possible to evaluate the performance of the generative model using existing metrics and datasets at hand.
\end{enumerate}

Due to the novel nature of our research question, which will be further discussed in Section \ref{sec:related-work}, this project is an exploratory study. Our goal is to evaluate the possibilities in this field and assess the feasibility of text-conditioned art generation, rather than developing a comprehensive solution to our research questions.

\section{Relationship to Prior Work}
\label{sec:related-work}
Our work is relevant to two main subjects of literature: conditional music generation and multimodal machine learning. Inspired by recent advancements in natural language generation and image generation with deep learning, there has been an increasing interest in applying deep neural networks to music generation. Popular model architectures include variational autoencoders (VAE), long short-term memory networks (LSTM), and Transformers. 

MusicVAE \cite{pmlr-v80-roberts18a} uses an LSTM encoder coupled with a hierarchical LSTM decoder to learn a latent space representation of the input over variational inference loss. The latent embedding it learns allows it to perform style-conditioned music generation and interpolation between two pieces. Music Transformer \cite{DBLP:journals/corr/abs-1809-04281} is one of the first works that use a Transformer decoder to generate minute-long compositions. To enforce the model to generate with coherent structure, Theme Transformer \cite{shih2021theme} devises a procedure to train the model to generate music that contains several repetitions (with possible variations) of a specified theme sequence. Authors report improvements across all metrics compared with Music Transformer. One of the latest works, MuseMorphose \cite{DBLP:journals/corr/abs-2105-04090}, is an encoder-decoder Transformer model trained with variational inference. It employs cross-attention between encoder and decoder to generate coherent music pieces. While these previous works are able to generate music conditioned on certain style inputs or seed sequences, none of them are trained on multimodal data, thus they are not able to generate music conditioned on natural language inputs. 

The second research area that our work is related to is multimodal machine learning. CLIP \cite{Radford2021LearningTV} is a multimodal language and vision model that is trained on several self supervised pretraining objectives to learn whether a given piece of text is related to a given image. This is achieved by training through a contrastive loss where the training data is obtained from internet scraping of 1.28 Million images and their captions. The contrastive objective function constraints the similarity metric between positive pair (related image and caption) be higher than a negative pair (unrelated image and caption). Various recent works, including \cite{Brock} and \cite{smith2021clip}, have demonstrated that CLIP could be used to train text conditioned generative image models. However, while the alignment between computer vision and natural language has become a popular area of discussion thanks to CLIP, music remains as a relatively less-studied modality. In fact, to our best knowledge, no work has been done in the field of multimodal machine learning of music and natural language. 

In our work, we would like to bring together the two and develop a text-conditioned music generation model using a CLIP-like text-music alignment model and a conditional music generation model.

\section{Project Requirements}

\subsection{Intended Users}

Stakeholders of this project can mainly be categorized into two types: machine learning researchers and multimedia artists. The most common use cases for machine learning researchers include: applying our model to related tasks, using our model as a baseline, improving our model through better design or training procedures, and extending our model by further explorations. We provide direct access to our source code and pretrained models for these purposes. 

Multimedia artists can use our end-to-end model for music generation, as well as other types of art generations. Our model pipeline can be easily accessed by running the bash script we provided, in addition, our pretrained checkpoints will also be available at HuggingFace for easier usages for non-technical audiences.

\subsection{System Functionality}

The major functions of our system are as follows:
\begin{itemize}
    \item \textbf{F1. Representation Learning: }Our system will include an encoder that learns the latent embedding of an inputted music piece.
    \item \textbf{F2. Text Conditioned Music Generation:} Our system is a generative model that produces music conditioned on text prompt.
\end{itemize}

\subsection{Non-Functional Requirements}

We have multiple components in our model: a CLIP-based encoder that converts text to a latent representation space from which music be generated, a music decoder that generate music based on the latent representation, and a discriminator to evaluate the quality of the music.

We expect the latency during inference to be around 10-15 minutes. This is because each time a piece of music is to be generated, the music decoder’s parameters are updated based on the loss function as assessed by music CLIP. This process continues until the generated music is faithful to the input text prompt. 

The second constraint we have is the non-deterministic nature of model outputs. As one of the broad intentions of the project is to enable creative artists to generate and explore with these models, it is important to realize that the same set parameters might yield different music each time the model is used, which is different from the traditional music generation process.

\subsection{Resource Requirements}

\subsubsection{Data Requirements}
Our greatest challenge is the data requirements. We require a sufficiently large music-text dataset to train our CLIP-based text-music alignment model. Besides, we require the music data to be piano-based music expressed in a symbolic format that could be converted to the format our model expects.

\subsubsection{Computing Requirements}
We expect the model training process to be extremely computational-heavy, especially if the dataset is larger. As a reference, the best performing vision Transformer used by CLIP took 12 days to train on 256 V100 CPUs \cite{Radford2021LearningTV}.


\section{Experiment/System Design Overview}

Our machine learning system is designed to consist of the following four components:
\begin{enumerate}
    \item Data Preprocessing
    \item Feature Engineering
    \item Model Training
    \item Model Evaluation
\end{enumerate}

\begin{figure}
    \begin{center}
        \includegraphics[width=0.95\textwidth]{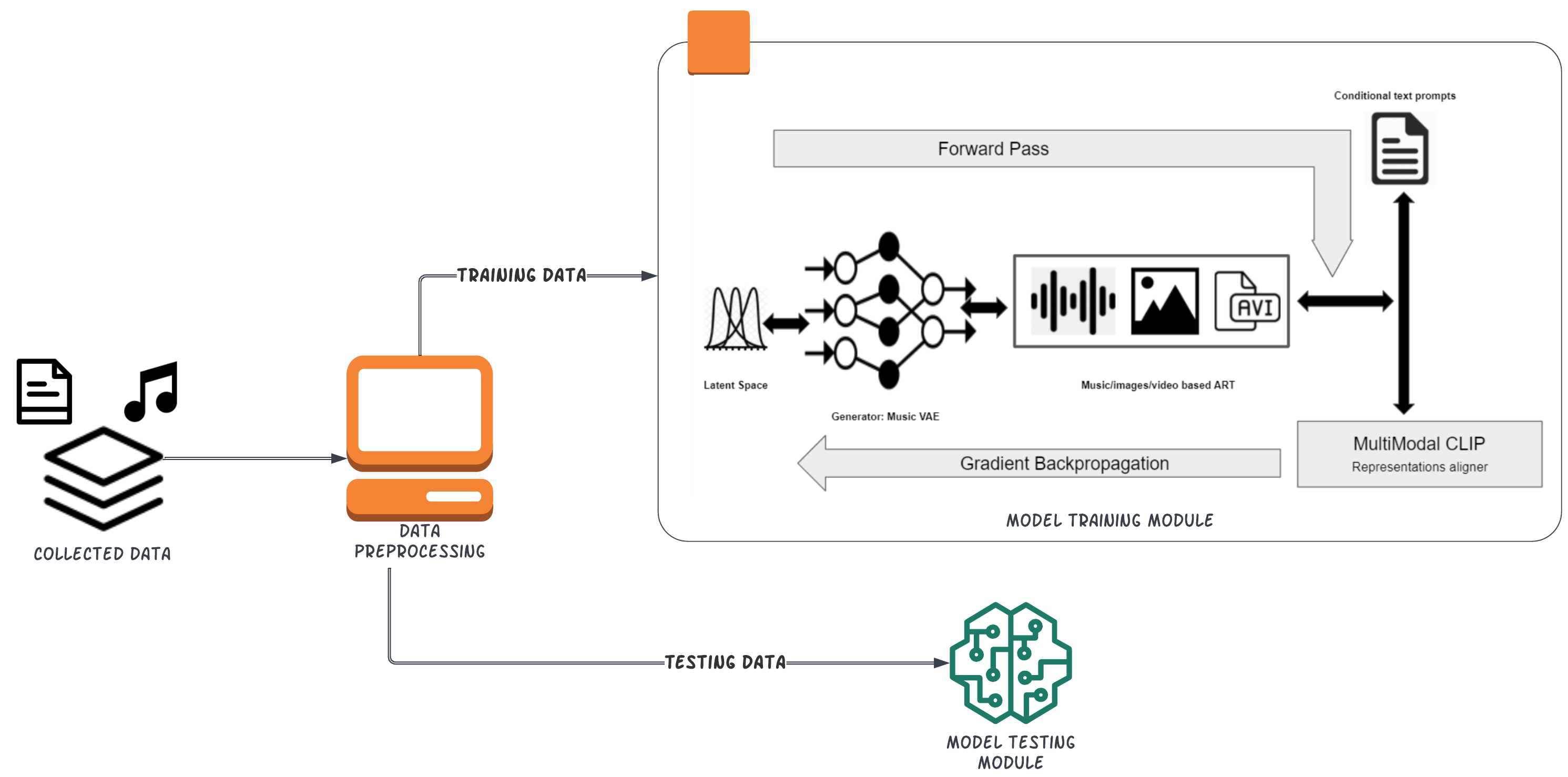}
        \caption{System Design Diagram}
        \label{fig:system-design}
    \end{center}
\end{figure}

Figure \ref{fig:system-design} shows an overall workflow diagram that displays how our system is intended to operate at the module level. To start with, collected data are passed into the data pre-processing module, where datasets are filtered, matched, and integrated. This module also includes the feature engineering component, in which music data are converted from their original format into our designed representation (see details below). After the data pre-processing module, a train-test split is performed on the resulting dataset. The training data is passed into the model training module, and the testing data will be reserved for the model evaluation phase.

The subsections below introduce each module in more detail.

\subsection{Data Preprocessing}
\begin{figure}[h!]
    \begin{center}
        \includegraphics[width=1\textwidth]{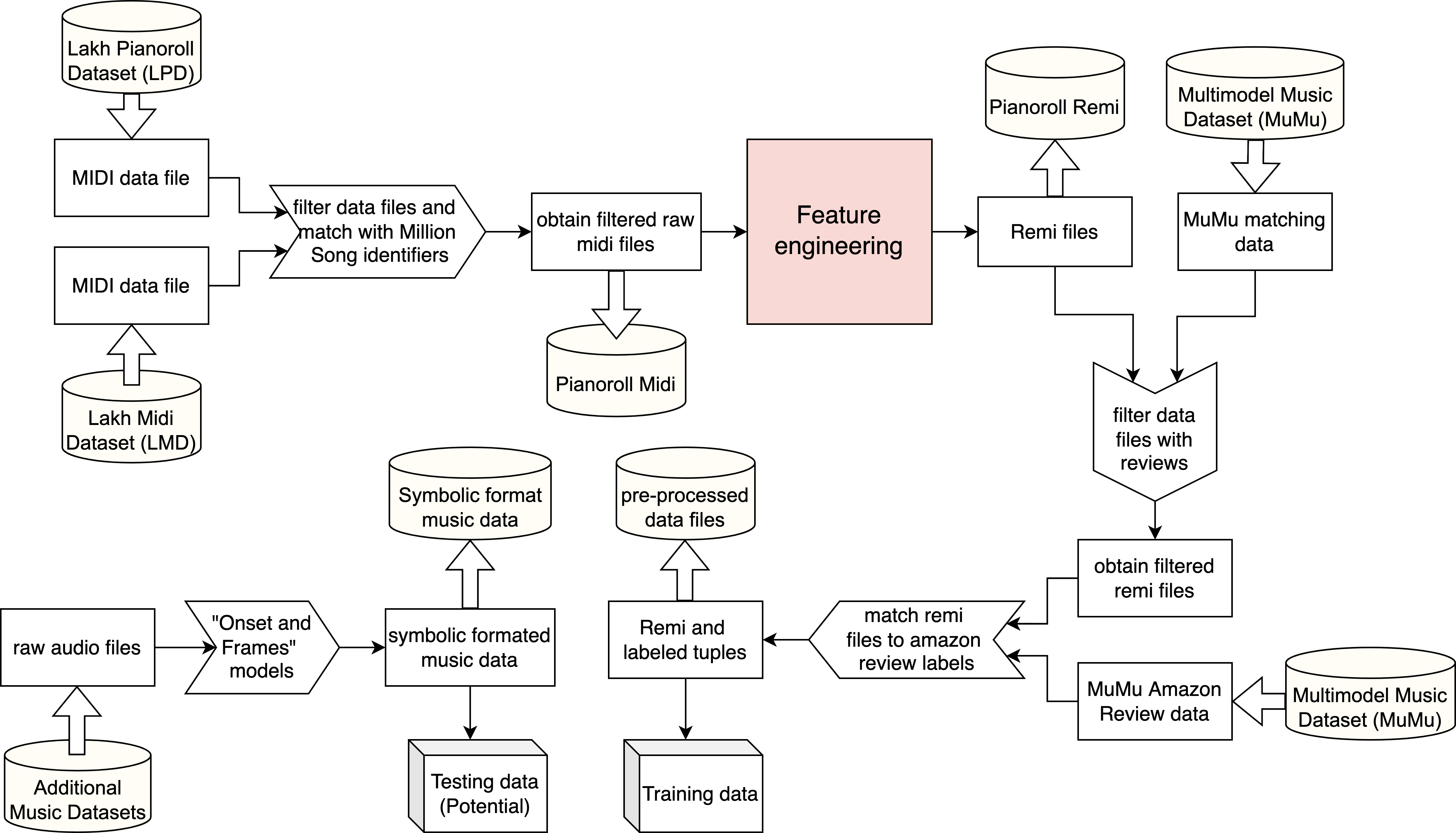}
        \caption{Data Design Diagram}
        \label{fig:data-design}
    \end{center}
\end{figure}

Figure \ref{fig:data-design} shows an overview of the data processing procedures. We used the Lakh Midi Dataset (LMD) \cite{LMD} as starter raw MIDI files. This dataset consists of symbolic formatted music data that could be matched with files from the Million Song Dataset (MSD) \cite{MSD}. We use the Lakh Pianoroll Dataset (LPD) \cite{LPD} as target files to filter the raw MIDI files. LPD contains a list of music files of various genres without lyrical content. This filtered dataset is fed into the feature engineering process.

The output of feature engineering is a set of REMI files. We use a list of matching data provided in the Multimodel Music Dataset (MuMu) \cite{MuMu} to obtain a list of REMI files with MSD identifiers that are able to match Amazon review data. The processed Amazon Reviews from MuMu are then matched with the REMI files as (data, label) tuples for the training process.

We may potentially include additional raw music datasets for comparison and testing purposes if needed. These datasets would have to apply “Onsets and Frames” models \cite{OnsiteFrames} to convert the raw audio data into a symbolic format.

\subsection{Feature Engineering} \label{feat-eng}

\begin{figure}[h!]
    \begin{center}
        \includegraphics[width=1\textwidth]{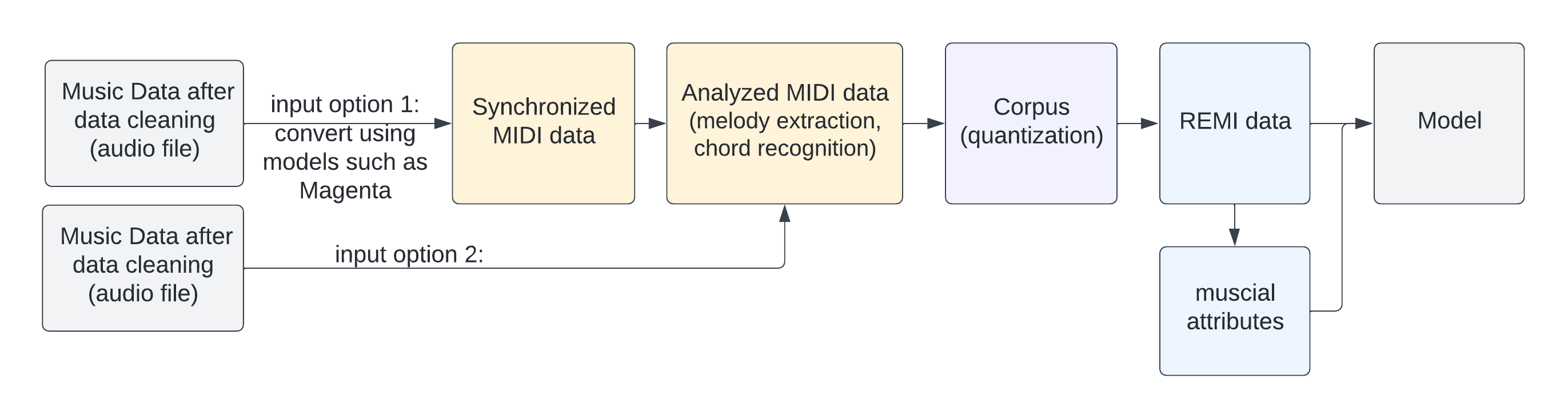}
        \caption{Feature Engineering Diagram}
        \label{fig:feat-eng}
    \end{center}
\end{figure}

Figure \ref{fig:feat-eng} provides a general overview of the feature engineering process connecting the output of the data flow process and the input of the model training module. 

 The starting point, namely the output of the data flow process, of our feature engineering will be data of audio format or MIDI format. Currently, we are using LMD, which is a dataset in MIDI format, but later on, we might add additional data in either MIDI format or in audio format. 
 
 We will be mainly using revamped MIDI derived events (REMI) as our music representation \cite{REMI}. We follow similar steps as Hsiao et al. did in their work to generate the REMI data \cite{compound2021}. If the data are audio files, we will convert them into synchronized MIDI data and then further extract melody and chords. A corpus is then built based on the analyzed MIDI data and is transformed into REMI format. We follow the input format used by MuseMorphose \cite{MuseMorphose}, so we further compute the musical attributes, raw rhythmic intensity scores and raw polyphony scores as defined in section \ref{dataset-section}. We divide the attribute scores equally into 8 bins to obtain the \emph{rhythmic intensity} class ($a
^\textnormal{rhym}$) and the \emph{polyphony} class ($a^\textnormal{poly}$) \cite{MuseMorphose} from the generated REMI data and feed the concatenated results into the model.

\subsection{Model Training}
In this section we describe various components of the model training architecture we follow for both the text to music generation.

\subsubsection{Conditional Multimodal Generative Architecture}
\label{core model}
\begin{figure}[h!]
    \begin{center}
        \includegraphics[width=1\textwidth]{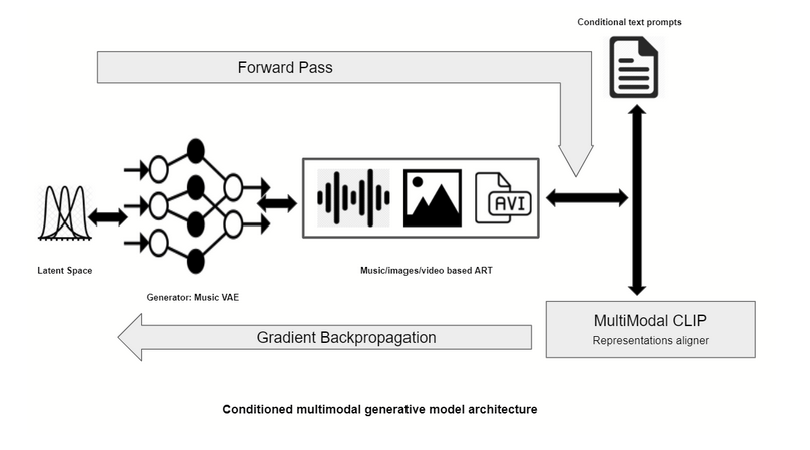}
        \caption{Conditional MultiModal Architecture}
        \label{fig:cond-mm-arch}
    \end{center}
\end{figure}

The figure \ref{fig:cond-mm-arch} describes the overall end-to-end model flow chart. The model architecture has three main components: The individual modality encoders, cross modality encoder, and the CLIP model to assess the faithfulness of the generated music with respect to text.

\subsubsection{Contrastive Loss Based Multimodal CLIP }

\begin{figure}[h!]
    \begin{center}
        \includegraphics[width=1\textwidth]{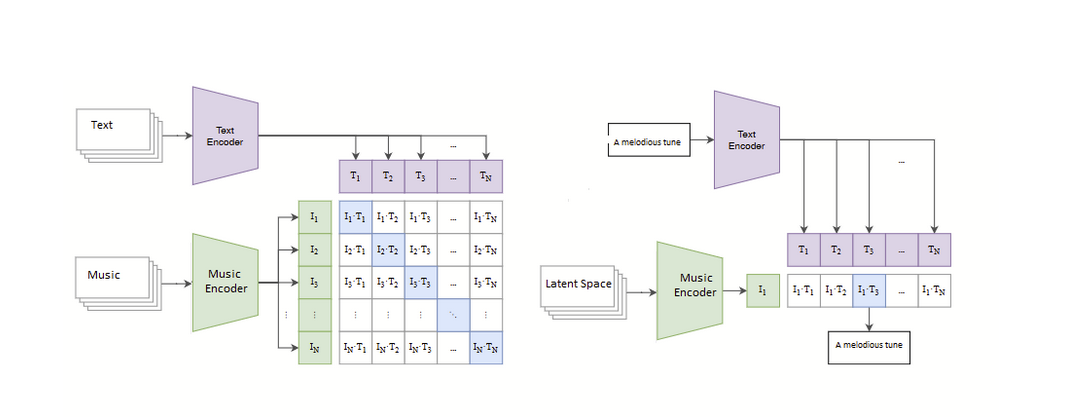}
        \caption{Contrastive loss based multimodal CLIP}
        \label{fig:music-clip}
    \end{center}
\end{figure}

In our project on conditional multimodal art generation, one of our primary focuses is generating music from text. To do so, we will be training a model that is similar to CLIP in architecture that learns the alignment between music and text. To generate music, we would be using MusicVAE. Broadly, we have a generator module, evaluator module. The errors of the evaluator module evaluates the generated music and it's relation to input condition. These errors are backpropogated till the input latent space for the musicVAE. The parameters of the MusicVAE would be updated in such a way that it learns to please the evaluator module. Figure \ref{fig:music-clip} describes the outline of the training and inference of the multimodal model.

It is important to realize that, the entire generator model has to go through a few (~100) rounds of gradient updates for any given text prompt. This is typically a time consuming process (resulting in ~10 min latency to generate aligned music with a given piece of textual context). This latency is due to the sizes of the generators with millions of parameters that need to be updated until the CLIP model's assessment of the art is in aligned with the input.

Contrastive loss between the music and the text encoders is used during the error back propagation. 
 
\subsubsection{Cross Modal Encoder}
\begin{figure}[h!]
    \begin{center}
        \includegraphics[width=1\textwidth]{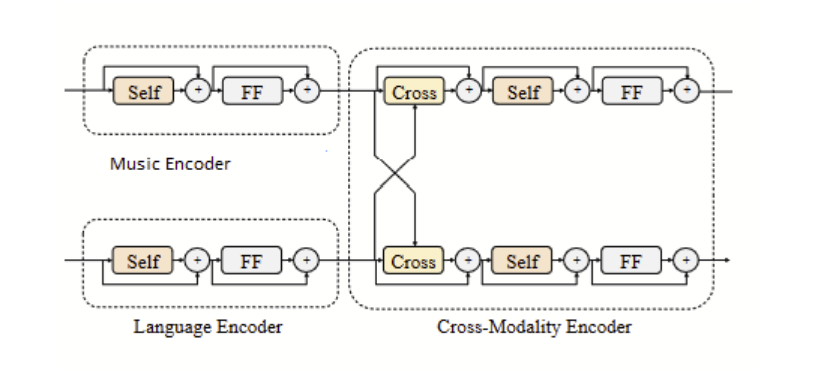}
        \caption{Cross Modal Attention}
        \label{fig:crossmodal-attn}
    \end{center}
\end{figure}

This section describes how the cross modal encoder is constructed. As described in the figure \ref{fig:crossmodal-attn} The basic building blocks of the models are self attention based transformers. Initially, the text and music pass through their corresponding encoders. The resulting representations are passed into a cross modal attention block where the query vectors of music (language) are attended to the key vectors of language (music) to evaluate the relationship between the various tokens of the each other modalities. 

\subsection{Model Evaluation}
\label{sec:model-evaluation}
\begin{figure}[h!]
    \begin{center}
        \includegraphics[width=1\textwidth]{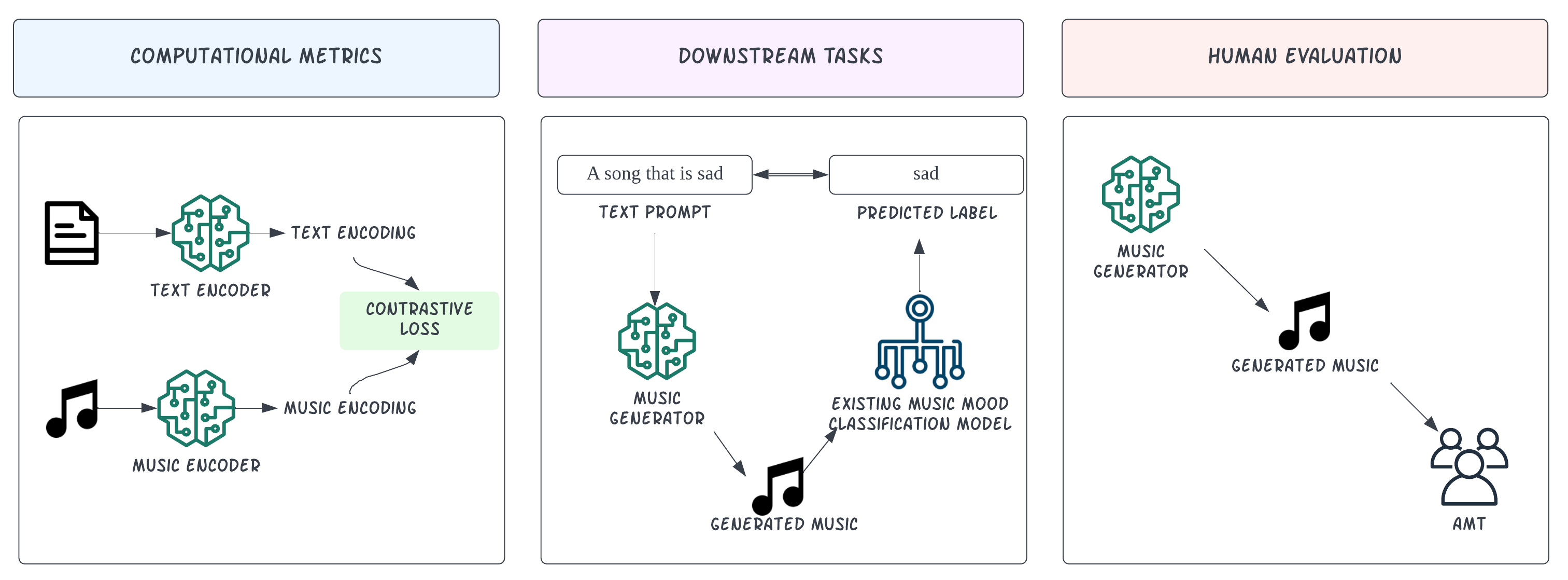}
        \caption{Generative Model Evaluation Methods}
        \label{fig:model-testing}
    \end{center}
\end{figure}

Figure \ref{fig:model-testing} displays three common evaluation methods for generative models: objective metrics, downstream tasks, and human evaluation.\newline 
In the second method, downstream tasks such as music mood classification will be used to indirectly test the performance of the music generation model. We could ask the music decoder to generate a piece of music based on a text prompt (e.g. “a song that is sad”) and then ask a pre-trained mood classifier to classify the emotion of the piece. However, given the limited time we have, we have decided to prioritize the first and third evaluation methods, which are more straightforward and easier to implement: 
\begin{itemize}
    \item \textbf{Computational Metrics:} Since we are training the alignment model using contrastive learning, how well the model is learning will be reflected in its train and validation loss. We look at loss curves during training to obtain a basic sense of how well the model is learning. Besides, we have also implemented 7 objective evaluation metrics to measure the quality and diversity of the generated music. They fall into three categories:
    \begin{itemize}
        \item Integrity-related metrics measure the general quality of the music:
        \begin{itemize}
            \item Qualitified notes rate
            \item Empty bar rate
        \end{itemize}
        \item Pitch-related metrics include summary statistics of music pitches and reflect the range and diversity of the music:
        \begin{itemize}
            \item Min, max, and space of pitches
            \item Unique pitches per bar
        \end{itemize}
        \item Rhythm and harmony related metrics reflect the rhythmic and harmonic complexity of the music:
        \begin{itemize}
            \item Chord repetition
            \item Polyphonicity 
            \item Rhythmic intensity 
        \end{itemize}
    \end{itemize}
    \item \textbf{Human Evaluation:} The most direct and accurate method to evaluate a generative model's performance is through human evaluation. We will listen to model outputs and invite fellow students and friends as listeners to rate the quality of the generated music based on its coherency and consistency on a Likert scale. 
\end{itemize}

\section{Experimental Design}
Our experiments are designed based on our downstream tasks. We will be running our model on music generation tasks given natural language inputs, evaluated by human evaluation scores, including but not limited to coherency scores and consistency scores. We predict that our model will yield better results than unimodal generators. 

\subsection{Dataset} \label{dataset-section}
As explained in section 6.1, we used an external dataset MuMu to match the processed REMI files with Amazon reviews with an Amazon identifier field defined by MuMu. Table \ref{dataset stats} displays summary statistics of these datasets. Multiple recordings may be derived from the same track with small variations in music. 

\begin{table}[h!]
  
  \centering
  \begin{tabular}{lll}
    \toprule
    Name     & Lakh Midi     & Lakh Pianoroll \\
    \midrule
    \# of unique tracks  & NA  & 21,425   \\
    \# of unique recordings     & 176,581 & 174,154      \\
    \# of matches with MSD & 45,129       & 115,160  \\
    \bottomrule
    
  \end{tabular}
  
  \caption{Summary statistics of LPD, LMD, and MuMu}
    \label{dataset stats}
\end{table}

The MuMu dataset consists of 95,6292 Amazon reviews. It matches MSD track identifiers with Amazon identifiers, and each MSD track may have multiple review text labels. Table \ref{matching-stats} shows a summary of these numbers. 
We use unique recordings from LPD to match different Amazon review text labels in MuMu, and segment each recording to equal lengths of 16 bars. Each music segment is matched with both its corresponding amazon review as positive labels, as well as a random incorrect amazon review as negative labels. The output files will be of type (String, String) tuple, where the first is REMI objects converted to strings, and the second is Amazon review text in string format. 
We have a total of 684,004 positive music and label tuples and an equal number of negative tuples available. We divide this into 0.8/0.1/0.1 for training/validation/testing splits.

\begin{table}[h!]
  
  \centering
  \begin{tabular}{ll}
    \toprule
    \# of tracks with matching Amazon labels     & 2,333 \\
    \# of recordings with matching Amazon labels  & 11,697     \\
    \# of Amazon labels matched to track identifiers & 76,861 \\
    \bottomrule
    
  \end{tabular}
  
  \caption{Summary stats of relationship between LPD and MuMu}
    \label{matching-stats}
\end{table}


\begin{table}[h!]
  
  \centering
  \begin{tabular}{lll}
    \toprule
    Event Type     & Description     & \# tokens \\
    \midrule
    \textsc{Bar}  & beginning of a new bar  & 1  \\
    \textsc{Sub-beat}   & position in a bar, in 16th note steps & 16      \\
    \textsc{Tempo} & 32$\sim$224 bpm, in steps of 3 or 6 bpm    & 65 \\
    \textsc{Pitch} & MIDI note numbers (pitch) 0$\sim$127    & 128  \\
    \textsc{Velocity} & MIDI VELOCITIES 40$\sim$126, in steps of 2   &44  \\
    \textsc{Duration} & multiples (0$\sim$16 times) of 16th note      & 17 \\
    \textsc{Chord} & chord markings (root \& quality)    & 133 \\
    \textsc{Eos} & end of sentence    & 1 \\
    \midrule
    \textbf{All events} & \textemdash & \textbf{405}\\
    \bottomrule
  \end{tabular}
  
  \caption{The vocabulary used to represent the dataset.}
    \label{remi-events}
\end{table}

Table \ref{remi-events} shows the event definition used to represent the dataset after feature engineering. As is mentioned in section \ref{feat-eng}, we follow the same definition used by MuseMorphose \cite{MuseMorphose}. 

In addition to the REMI events, we also calculate the musical attributes, rhythmic intensity and polyphony, as factors that determine musical emotion \cite{MuseMorphose}. The rhythmic intensity score, $s^\textnormal{rhym}$, measures the percentage of sub-beats with at least one note onset \cite{MuseMorphose}. The polyphony score, $s^\textnormal{poly}$, is defined as the average number of nodes being \textit{hit} (onset) or \textit{held} (not yet released) in a sub-beat \cite{MuseMorphose}. 





\subsection{Machine Learning Models/Algorithms}

For the text to music alignment, we will be starting with Open AI CLIP model \cite{Radford2021LearningTV}. We will replace the image encoder with the encoder of MuseMorphose \cite{MuseMorphose}. The training process starts with the pre-trained weights of the CLIP model. During inference, we will be using the decoder of MuseMorphose and the decoder acts as generator whose weights are to be updated as guided by the CLIP for music model.


Most of the earlier work in music encoding are decoding are done through RNNs \cite{pmlr-v80-roberts18a}. Recent works like the Theme transformer \cite{theme-transformer} use self attention-based transformers as the building blocks of the encoder-decoder architecture. However, theme transformer represents the music through customized events and is not easily generalized to all forms of art. Hence, we chose to use MuseMorphose \cite{MuseMorphose} as it uses a simple modified REMI format and the architecture is based on transformers. This will enable us to construct the cross-modal attention in the CLIP for the music model without any extra processing.

\subsection{Evaluation Metrics}
As explained in Section \ref{sec:model-evaluation}, our main evaluation metrics include objective metrics and human evaluation.
\begin{itemize}
    \item \textbf{Contrastive loss on the test data:} Given a (text, music) pair from the test data, contrastive loss minimizes the embedding distance when they are aligned and maximizes the distance otherwise. We use InfoNCE loss as our contrastive loss implementation. 
    \item \textbf{Music metrics:} We implemented 7 music metrics to measure the integrity, diversity, and complexity of the generated music. Refer to Section \ref{sec:model-evaluation} for a complete list of our metrics. 
    \item \textbf{Human evaluation score:} Human listeners will be asked to rate the generated music on a Likert scale from 1-5. The aspects of evaluation are coherency (i.e. how coherent and smooth the music sounds by itself) and consistency (i.e. how well the music matches the input text description). 
\end{itemize}

\section{Test Design}

As described in Section \ref{sec:model-evaluation}, we test our model in two ways using the following procedures:

\begin{itemize}
    \item \textbf{Training Loss:} We record the training and testing contrastive loss with respect to the number of training epochs. See Section \ref{sec:result} for details. While the current figure only includes the values of the baseline model, we expect the final results to include the loss curves of various experiments using different hyperparameters. 
    \item \textbf{Computational Metrics:} We evaluate the generated music on the objective metrics descried in Section \ref{sec:model-evaluation}. 
    \item \textbf{Human Evaluation:} We listen to pieces generated by our model and rate outputs on a Likert scale.
\end{itemize}

\section{Deployment Model}
This project aims to explore deep learning methodologies for music generation. Thus we are not expecting any deployments for this project. We will be delivering our results in Jupyter Notebook format with documentation of our code, as well as a report for our findings and results. 
We may also consider a web UI for displaying our results and pre-trained model for public use. This user interface will contain sliders to control attributes of the generated music, where users can select music based on preferences. We will deploy this UI using Streamlit.

\section{Risks/Challenges}

In this section, we detail the potential risks we may face, the challenges we will have to address and potential solution framework to address these issues.

\subsection{Quality of Art}
It is important to realize that the generated pieces of art won’t be perfect at the onset. We need to keep tuning various model hyper parameters till the generated art is well aligned with the expectations. This requires an artistic effort by a human creator. An artist playing on a piano need not understand the intricacies of constructional and destruction interference of sound waves over a string (that results in the generation of sounds of various notes and pitches. Similarly, in our case, the artist need not understand the model or the mathematical abstractions therefore. One need to solely experiment with and tune the parameters till the quality of art is perfect.


\subsection{Inference Times}
Unlike the typical workflow of any machine learning model, in our case, during the inference, the generator is retrained as guided by the CLIP based model. This involves updating several millions of parameters (of the musemorphose music decoder) through backpropogation at the time of inference. Approximately, each epoch of decoder training during the inference takes 4-5 min. However, music typically needs a smooth pattern across bar. We would have to resort to using these patterns to generate a set of consecutive bars without having to retrain the generator for each bar

\section{Tools and Dependencies}

The tools and libraries that we use include but are not limited to the following:
\begin{itemize}
    \item \textbf{Python3:} Most of the libraries we ought to use are developed in Python3.
    \item \textbf{Magenta:} Magenta\footnote{https://github.com/magenta/magenta} is a set of models developed by Google Magenta team, available as a pip package. Our project will most likely reference some Magenta models, e.g. MusicVAE, Music Transformer, etc.
    \item \textbf{python-rtmidi:} \footnote{https://pypi.org/project/python-rtmidi/}We will need this Python library to handle realtime MIDI input and output.
    \item \textbf{MuseMorphose:}\footnote{https://github.com/YatingMusic/MuseMorphose} Our implementation of music encoder and decoder are based on MuseMorphose.
\end{itemize}

\section{Results}
\label{sec:result}
\subsection{Data Pre-processing}
\begin{figure}[h!]
    \begin{center}
        \includegraphics[width=1\textwidth]{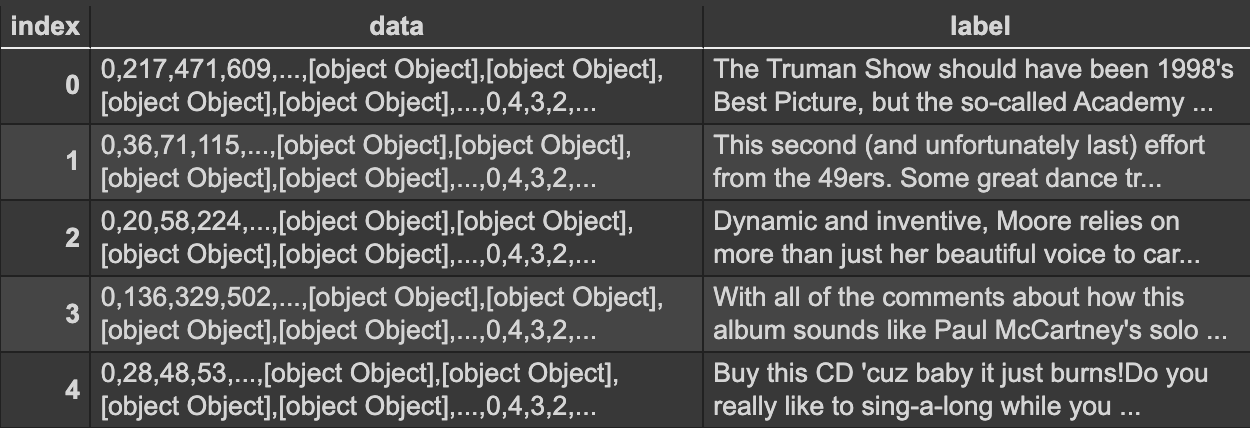}
        \caption{First 5 rows of the track-level training dataset}
        \label{fig:sample 5 rows}
    \end{center}
\end{figure}

We match all unique MSD tracks with the Amazon review dataset to obtain our track-level training dataset. The obtained dataset contains 2194 entries after excluding 7 broken files and 132 filtered files. A sample of the first 5 rows of the dataset is shown in Figure \ref{fig:sample 5 rows}.
Table \ref{dataset stats} shows some statistics of this dataset. While the shortest music piece contains just 8 events, the longest piece contains 47,357 events. A similar pattern could be observed in the Amazon review text data, which has a minimum length of only 1 token and a maximum length of 1,774 tokens. The large variance of data and label lengths in the sample data suggests a high level of heterogeneity in our data. Thus we segmented each piece into smaller chunks, having a total of 684,004 positive music and label tuples and an equal number of negative tuples.

\begin{table}[h!]
  \centering
  \begin{tabular}{llll}
    \toprule
    Name     & Min     & Max & Average \\
    \midrule
    \# of events in Remi formatted data & 8 & 47357 & 4029.75   \\
    \# of tokens in Amazon review labels     & 1 & 1774 & 120.21      \\
    \bottomrule
  \end{tabular}
  \caption{Summary statistics of sample processed data}
    \label{dataset stats}
\end{table}

\subsection{Feature Engineering}
Preprocessed data are transformed into four features: bar position, REMI events, rhythmic intensity class, and polyphony class. Bar position is represented by a list of integers that are indices of the start of a bar. 

The REMI event items are in the format of dictionary having `name' and `value' as their keys. The `name' corresponds to the event types defined in table \ref{remi-events}. An example of the REMI data items is shown below:
\begin{itemize}
    \item \{`name': `Bar', `value': None\}
    \item \{`name': `Beat', `value': 0\}
    \item \{`name': `Chord', `value': `N\_N'\}
    \item \{`name': `Tempo', `value': 68\}
    \item \{`name': `Beat', `value': 12\}
    \item \{`name': `Note\_Pitch', `value': 64\}
    \item \{`name': `Note\_Velocity', `value': 72\}
    \item \{`name': `Note\_Duration', `value': 840\}
    \item \{`name': `Note\_Pitch', `value': 60\}
    \item \{`name': `Note\_Velocity', `value': 76\}
\end{itemize}

The music attributes, the rhythmic intensity class and the polyphony class, are obtained from the raw rhythmic intensity score and the raw polyphony score defined in section \ref{dataset-section}. After the raw bar-level scores are calculated, they are roughly equally divided into 8 bins and each bin is represented by an ordinal class from 0-7. An example of a list of rhythmic intensity classes can be [0, 3, 3, 3, 0, 5, 5, 5, 5, 5, 5, 6, 5, 5, 5], and an example of a list of polyphony classes can be [0, 3, 4, 2, 1, 6, 7, 7, 7, 7, 7, 7, 6].

\subsection{Training the Text-Music Alignment Model}
\begin{figure}[h!]
    \begin{center}
        \includegraphics[width=1\textwidth]{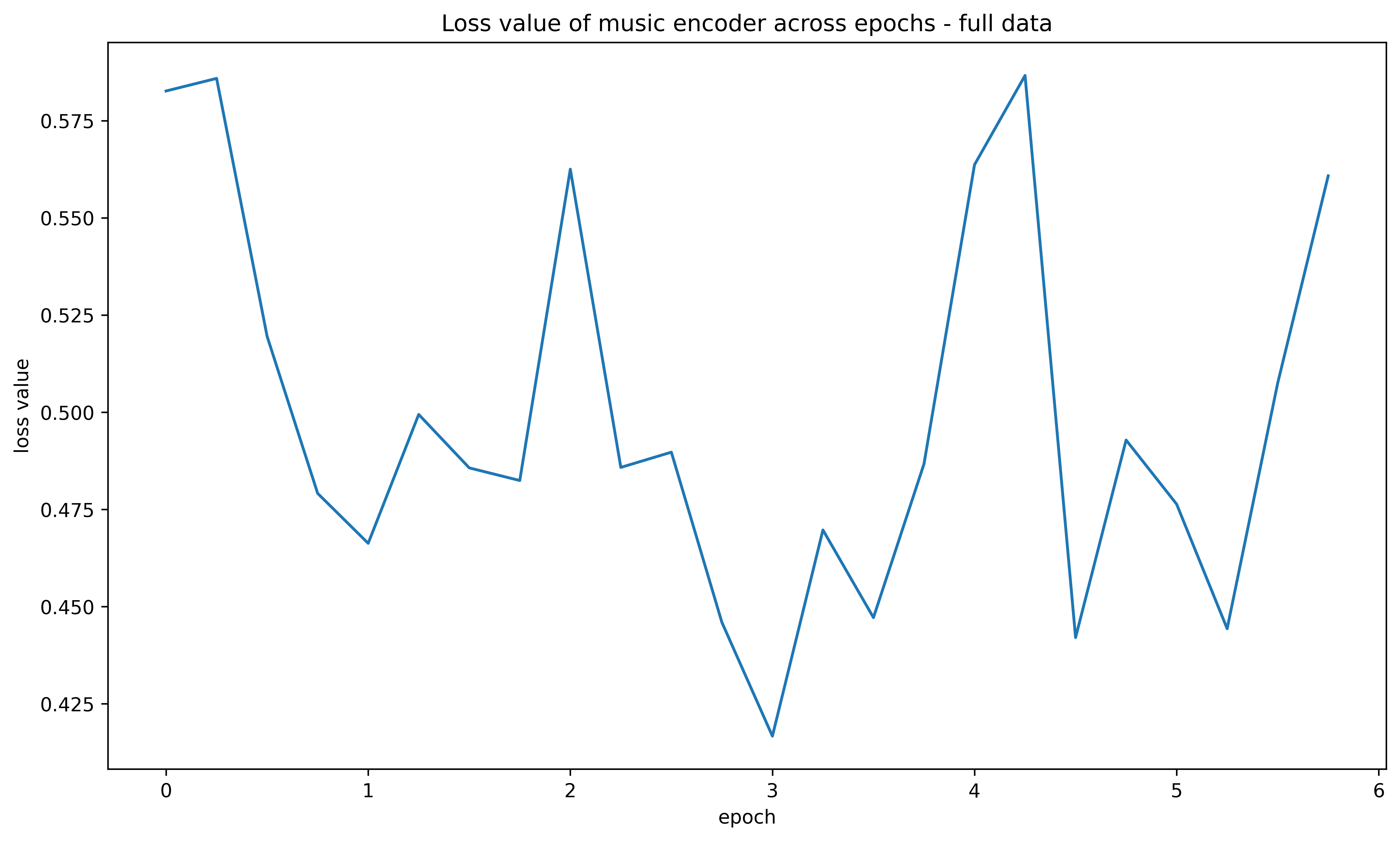}
        \caption{Plot of loss value on model trained with full data w.r.t. number of epochs trained}
        \label{fig:result-loss-full}
    \end{center}
\end{figure}

\begin{figure}[h!]
    \begin{center}
        \includegraphics[width=1\textwidth]{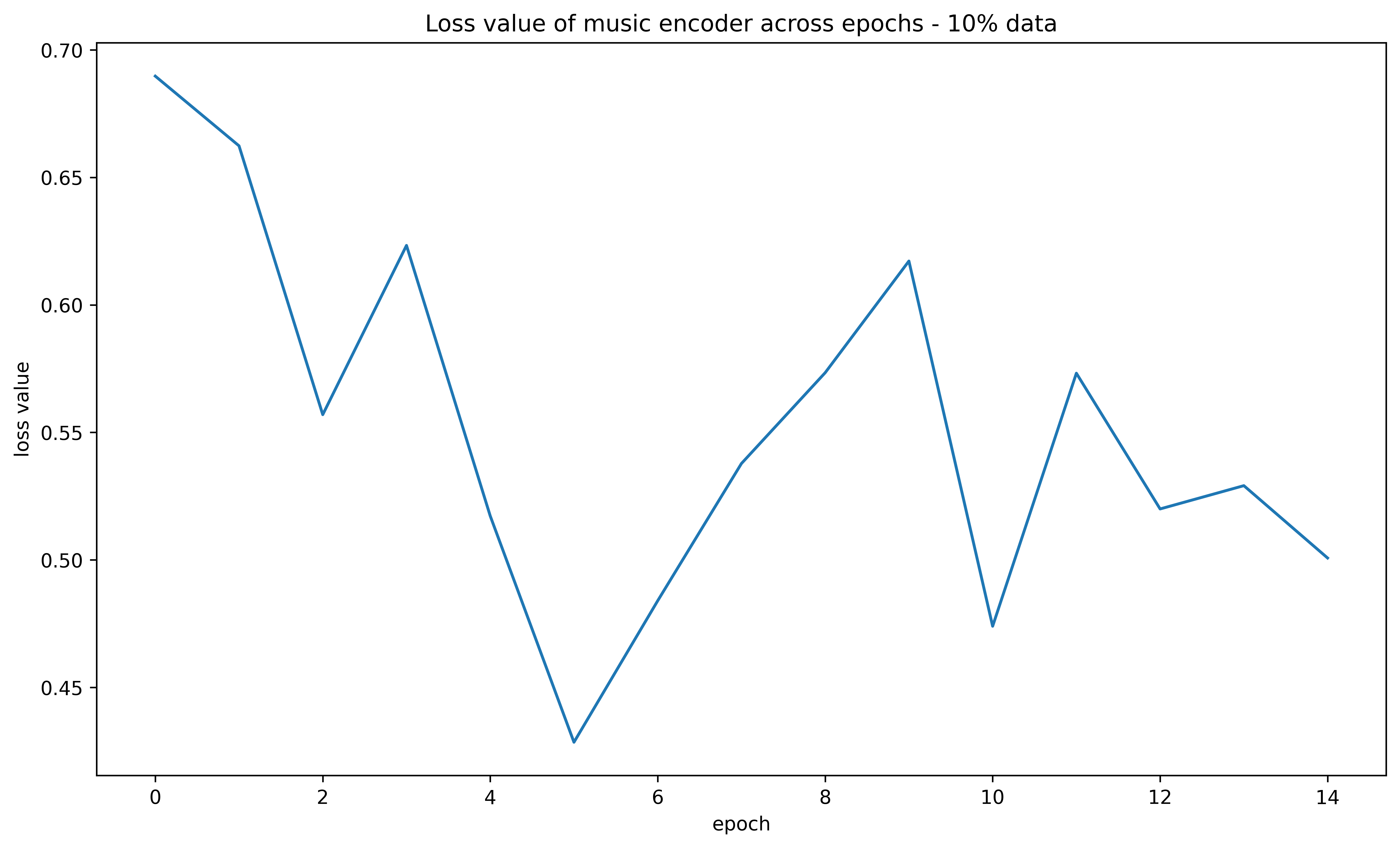}
        \caption{Plot of loss value on model trained with 10\% of data w.r.t. number of epochs trained}
        \label{fig:result-loss-baseline}
    \end{center}
\end{figure}

\begin{table}[h!]
  \centering
  \begin{tabular}{lllll}
    \toprule
    Data used     & batch size & lr & Min loss   & \#epochs to converge \\
    \midrule
    10\% data & 8 & 1e-4 &  0.4285 & 6\\
   full data & 8 & max: 1e-4, min: 5e-6, w/ scheduler & 0.4167 & 4 \\
    \bottomrule
  \end{tabular}
  \caption{Settings explored for encoder training}
    \label{encoder training results}
\end{table}

We performed two settings of encoder training experiments as shown in table \ref{encoder training results}. Our baseline is trained with 10\% of the total data, using a batch size of 8, where every epoch takes about 13k batches of data. We use a fixed learning rate of 1e-4. The training loss is plotted in figure \ref{fig:result-loss-baseline}. As shown in the plot, it takes 6 epochs for the baseline model to converge, and it reaches a minimum loss of 0.4285. Apart from the baseline model, we also trained the encoder on the full data with a batch size of 8, resulting in 137k batches per epoch. With a minimum learning rate of 5e-6 and a maximum learning rate of 1e-4, we applied a cosine annealing scheduler. The training loss for the model trained on full data is shown in figure \ref{fig:result-loss-full}. It takes 4 epochs for the model to converge, reaching a minimum loss of 0.4167. Training with full data converges faster in terms of number of epochs, but much slower in terms of running time. As is shown in figure \ref{fig:result-loss-baseline} and figure \ref{fig:result-loss-full}, training loss is fluctuating and being unstable, we hypothesize that this unstable training behavior is due to a suboptimal combination of our training parameters, namely the number of batches, learning rate, and learning rate scheduler.


\subsection{Training the Decoder During Inference}

\begin{figure}[h!]
    \begin{center}
        \includegraphics[width=1\textwidth]{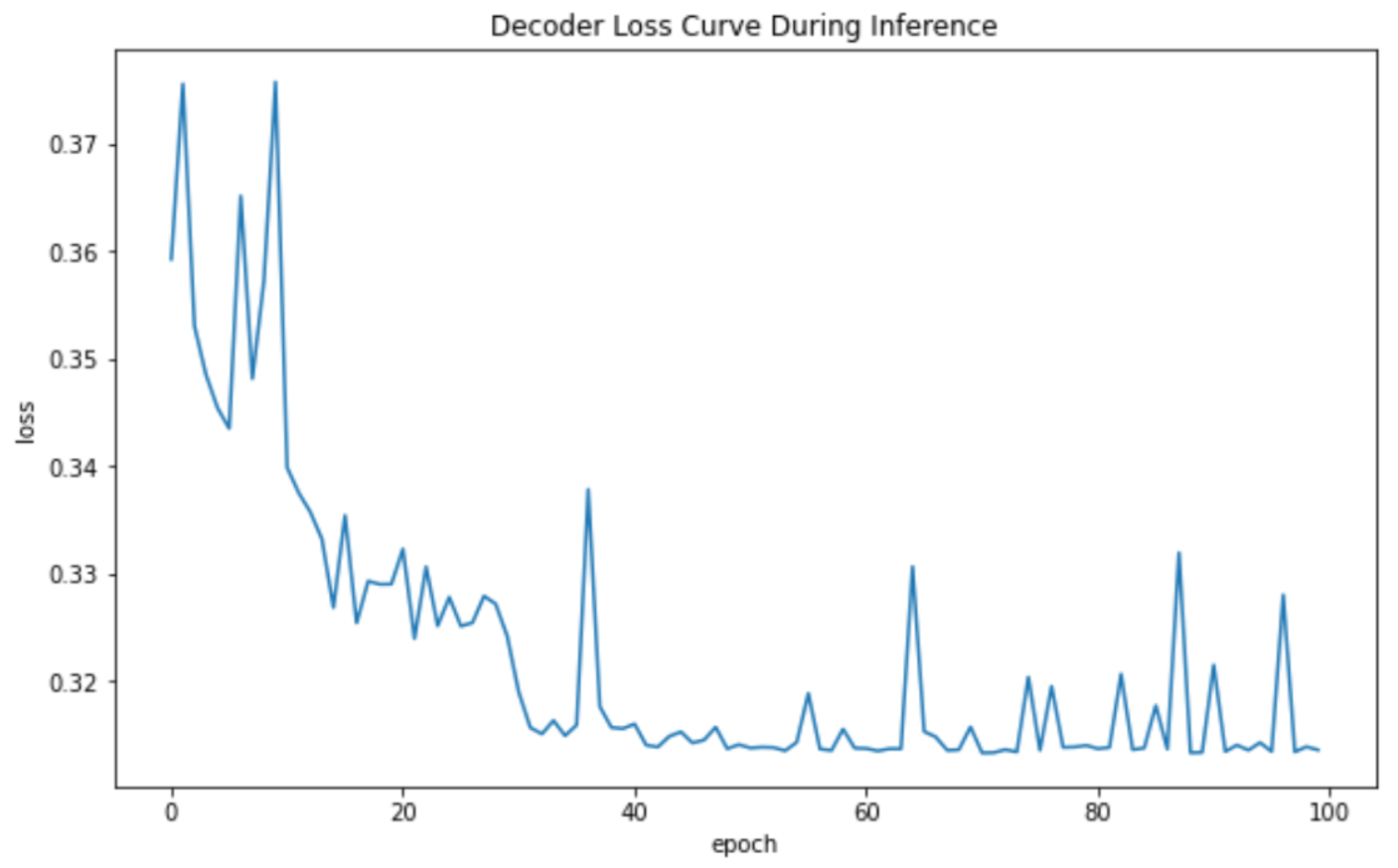}
        \caption{Loss curve of the decoder during inference. The decoder training converges before 100 epochs. }
        \label{fig:decoder-loss}
    \end{center}
\end{figure}

\begin{table}[h!]
\centering
\begin{tabular}{|l|l|}
\hline
Example 1           & Example 2           \\ \hline
Bar\_None           & Bar\_None           \\
Chord\_C\#\_sus4    & Tempo\_203          \\
Chord\_G\_m         & Chord\_C\#\_+       \\
Chord\_F\#\_m7      & Chord\_G\_o         \\
Note\_Pitch\_10     & Beat\_2             \\
Chord\_A\#\_7       & Note\_Pitch\_60     \\
Note\_Velocity\_110 & Note\_Duration\_240 \\
Note\_Pitch\_6      & Note\_Velocity\_66  \\
Note\_Velocity\_48  & Note\_Pitch\_80     \\
Tempo\_44           & Chord\_D\#\_m       \\
Note\_Velocity\_92  & Chord\_D\#\_o7      \\
Note\_Pitch\_32     & Note\_Pitch\_4      \\
Tempo\_194          & Chord\_B\_sus2      \\
Note\_Pitch\_73     & Tempo\_92           \\
Tempo\_104          & Note\_Pitch\_48     \\
Note\_Pitch\_43     & Note\_Pitch\_123    \\
Note\_Velocity\_108 & Note\_Pitch\_43     \\
......              & ......              \\ \hline
\end{tabular}
\caption{Example outputs from the decoder after training for 100 epochs. The outputs are in REMI representation. The first example was generated using the prompt "A pop song about love", the second example generated with the prompt "A song about using artificial intelligence to generate music for a class of graduate students".}
    \label{fig:decoder-outputs}
\end{table}

The decoder training during inference typically takes 50 - 60 epochs to converge, as shown in Figure \ref{fig:decoder-loss}. Figure \ref{fig:decoder-outputs} shows some example outputs from the decoder after 100 epochs of training. The outputs are in REMI representation, with each row representing an event token that was generated using nucleus sampling. 

Sample audio files that are converted from REMI representations are available on our Github repo \footnote{https://github.com/ZhouyaoXie/Intelligent-Multimedia-Art-Generation}. The samples were generated using three different prompts:
\begin{enumerate}
    \item A pop song about love
    \item A melodious tune for coping up with pain
    \item A song about using artificial intelligence to generate music for a class of graduate students
\end{enumerate}
Since the outputs contain both note events ("NOTE\_PITCH\_X" tokens) and chord events ("CHORD\_X" tokens), we assigned two different sounds to them to help distinguish them. We used an organ sound for the chords and a piano sound for the notes to generate these samples.

\section{Discussion and Error Analysis}

To summarize, our experimental process involves three major steps. First, we prepare the training data by matching MuMu with MSD and obtaining MIDI music data from LMD. We then convert MIDI data into modified REMI format and prepare data as input to the machine learning model. Next, we train our text-music alignment model on contrastive loss. Lastly, during the generation phase, we train the decoder on the same loss so that it learns to generate music that matches the text inputs.

\subsection{Data Prepartion}
One of the findings during the first step, data preparation, is that there does not exist a large text-MIDI multimodal dataset that we could utilize. Instead, we need to obtain matching text labels for music data. There are few available textual data available that could be matched with the LMD dataset. Our initial approach of using unique MSD tracks was only capable of producing 2,333 data tuples, which is insufficient for training purposes. Therefore, we used recordings in replacement, where multiple recordings may be mapped to the same track, and the same track id may also be able to map to multiple review texts. We performed inner join on the recording files and review data to obtain unique data and label tuples. This procedure improved the number of data files available for training from 2,333 to over 20,000.

In the context of the project’s vision, the lack of a text-MIDI multimodal dataset means that a lot remains to be done in the under-explored field of text-conditioned music generation. While we are able to obtain and process a dataset of size over 20,000, more work is needed to collect larger multimodal music datasets in the future.

\subsection{Encoder Training}

The training process involves learning the two encoders, one each for music, text, and the cross-attention modules. Each of these modules is a 12-layer transformer. We have used pre-trained weights for the BERT-based text encoder, and the Musemorphose-based music encoder. Given the millions of parameters involved, and as seen in our loss function behavior in \ref{fig:result-loss-full} \ref{fig:result-loss-baseline}, we need to increase the dataset size. However, each epoch takes around 2-3 days to finish with a single Tesla GPU. As we continue to improve the model performance, it is essential for us to use libraries to distribute the model training (not just the data parallelism) across multiple GPUs simultaneously.

\subsection{Decoder Inference}

Since we do not impose any constraints on the generation process, we observe a few issues with the direct raw outputs from the decoder, such as the ones shown in Table \ref{fig:decoder-outputs}. Some of the main limitations are:
\begin{itemize}
    \item The raw outputs contain an overly dense distribution of chord events, sometimes on the order of hundreds of chord tokens per bar.
    \item There is more than one tempo event per piece, causing the piece to have an inconsistent tempo. 
    \item In REMI representation, a note pitch event should always co-occur with a corresponding note duration and a note velocity event. However, in our outputs this is not always the case.
\end{itemize}

Currently, to resolve these issues, we introduce a few post-processing steps to convert REMI representation outputs to MIDI format. We sample the first chord after each bar and beat token and abandon the rest of the chord outputs. To deal with the issue of inconsistent tempos, we fix the first tempo as the tempo of the entire piece and ignore any following tempo tokens if there are any. Lastly, we keep all note pitch events and assign a default note duration and velocity to each pitch if they are not generated. However, rather than dealing with these issues through rule-based post-processing procedures, a better approach would be to try to solve these problems at a model level during training. We will describe a few ideas for model-level improvement in \ref{sec:future-work}

\section{Future Work}
\label{sec:future-work}
This is an exploratory project and we are trying to set new baselines for a not-so-well-established problem statement. We started with the idea of venturing into Style Transfer of music from one style to the other (After the text-based generation is achieved). Later we proposed a model architecture that could be used for both music and video generation from text. Overtime, dealing with the complexities involved with music generation, we changed our plan to focus solely on the music generation part. If one were to take up the current architecture we have for video generation, it's best to generate optical flow for the continuity of video. 



\subsection{Diffusion Based Models}
Recently, we have seen rise of diffusion based models that obtained SOTA level in text-to-image generation. The diffusion based models provide an easy alternative to attention based transformers and have proven to provide more consistent generative output than transformers. In the case of music generation, we are yet to evaluate the performance of diffusion based models for decoding music and learning the alignment between text and music.

\subsection{Better loss functions}
From the results of the generated music, we can observe that the output has more chord events than note events and several tempo events per bar of output. This is a direct result of modelless music generation. For a coherent music generation, we need to incorporate a few rules about music in the loss function. Since most rules can't be implemented as a continuously differentiable loss function, we need to deploy clever techniques as proxies for loss function: for example, using a regularizer for chord events, etc. 

\subsection{Multi-Instrumental Music Generation}
For the music generation part, the dataset we have been curating is based on piano tunes. We used the MuseMorphose encoder and its decoder for the generation task. However, the generated art would be constrained in the sense that it has only instruments. Potential future work could extend the current framework to music with multiple instruments.


\section{Conclusion}
In this paper, we present a multimodal music dataset that we generated by combining the LPD, LMD, and MuMu datasets. This dataset consists of over 684k pairs of positive and negative samples, and is designed specifically for contrastive learning. We then describe a text-conditioned music generation model that uses a CLIP-like architecture with pretrained music and text transformers. Through experiments, we demonstrate that it is possible to train a music-text alignment model using contrastive loss, and to train a decoder to generate music from text inputs. However, we also identify several limitations of our current approach that require further research to address.


\section{Acknowledgments}
We would like to thank Prof. Eric Nyberg, our mentor, for his invaluable guidance and advice.

\newpage
\section{Terminology, Definitions, Acronyms, and Abbreviations}

Please see a list of terminologies and abbreviations we use in Table \ref{term}.

\begin{table}[h]
\begin{tabular}{p{0.35\linewidth} | p{0.6\linewidth}}
\hline
\textbf{Term} & \textbf{Description}                                                                                           \\ \hline
Symbolic Music Representation &
  A score representation of music with an explicit encoding of notes or other musical events. Examples include piano-roll, MIDI, MusicXML, etc \\ \hline
MIDI          & A format for digitally store music using a set of events to denote how the music should be played              \\ \hline
REMI          & Revamped MIDI-derived events, a representation of MIDI data following the ways humans read and understand them \\ \hline
Modified-REMI & Music representation used by MuseMorphose \cite{MuseMorphose} \\ \hline 
Music Mood Classification &
A subfield of research in machine learning for signal processing that focuses on developing models to classify music into emotion categories \\ \hline
\end{tabular}
\caption{Terminology and Abbreviations}
\label{term}
\end{table}

%
\bibliographystyle{unsrt} 
\bibliography{sample}  





\end{document}